\documentclass[aps,prb,floatfix,twocolumn,showpacs,10pt]{revtex4-1}

\usepackage{epsfig,amsmath,amssymb}

\begin{document}

\title{Theory of Spin Relaxation in Two-Electron Lateral Coupled Si/SiGe Quantum Dots}
\author{Martin Raith$^1$, Peter Stano$^{2,3}$, and Jaroslav Fabian$^1$}
\affiliation{$^1$Institute for Theoretical Physics, University of Regensburg, D-93040 Regensburg, Germany\\
$^2$Institute of Physics, Slovak Academy of Sciences, 845 11 Bratislava, Slovakia\\
$^3$Department of Physics, University of Basel, Klingelberstrasse 82, 4056 Basel, Switzerland}

\vskip1.5truecm

\begin{abstract}
Highly accurate numerical results of phonon-induced two-electron spin relaxation in silicon double quantum dots are presented. The relaxation, enabled by spin-orbit coupling and the nuclei of $^{29}$Si (natural or purified abundance), are investigated for experimentally relevant parameters, the interdot coupling, the magnetic field magnitude and orientation, and the detuning. We calculate relaxation rates for zero and finite temperatures (100 mK), concluding that our findings for zero temperature remain qualitatively valid also for 100 mK. We confirm the same anisotropic switch of the axis of prolonged spin lifetime with varying detuning as recently predicted in GaAs. Conditions for possibly hyperfine-dominated relaxation are much more stringent in Si than in GaAs. For experimentally relevant regimes, the spin-orbit coupling, although weak, is the dominant contribution, yielding anisotropic relaxation rates of at least two order of magnitude lower than in GaAs.
\end{abstract}

\pacs{72.25.Rb, 03.67.Lx, 71.70.Ej, 73.21.La}

\maketitle

\section{Introduction}

Since the proposal of Loss and DiVincenzo,\cite{loss1998:PRA} electron spins in semiconductor quantum dots have been in the perpetual focus of research on spintronics.\cite{Wolf2001:S,zutic2004:RMP,fabian2007:APS} In GaAs based qubits, which are the state of the art, the essential gate operations\cite{nielsen,Barenco1995:PRA,loss1998:PRA} for quantum computation\cite{divincenzo2000:FP,Ladd2010:N} have been demonstrated.\cite{Hanson2007:RMP,Brunner2011:PRL,Taylor2007:PRB,Laird2007:PRL,Ladriere2008:NP,petta2005:S,Meunier2011:PRB,Koppens2006:N,Nowack2007:S,Foletti2009:NP} But GaAs possesses a serious handicap for coherent spin manipulations---the nuclear spins.\cite{Khaetskii2001:PRB,schliemann2003:JPCM} Controlling this source of decoherence is of major interest and an active field of research.\cite{Petta2008:PRL,rudner2007b:PRL,Imamoglu2003:PRL,Stepanenko2006:PRL,koppens2005:S,Foletti2009:NP,Ivo2009:NP}

An alternative to III-V semiconductors with inherent nuclear spins are systems composed of atoms without nuclear magnetic moment, such as Si and C.\cite{Balasubramanian2009:N,Ladd2005:PRB,Morton2011:N} Natural silicon consists of three isotopes: $^{28}$Si (92.2\%), $^{29}$Si (4.7\%), and $^{30}$Si (3.1\%).\cite{Ager2006:PSS} Hereof only $^{29}$Si has non-zero nuclear spin (I=1/2), and purification can further reduce its abundance down to 0.05\%.\cite{Sailer2009:RRL,Wild2012:APL} For this reason, silicon-based quantum dots have become the new focus of interest, and recent progress emphasizes their perspectives.\cite{Borselli2011:APL,Maune2012:N,Prance2012:PRL,Nordberg2009:PRB} Another advantage of silicon\cite{Morton2011:N,Ladd2010:N} over GaAs is a larger $g$ factor, which allows spin manipulations in smaller magnetic fields. On the other hand, device fabrication of silicon dots is more challenging,\cite{Eriksson2004:QIP} the spin-orbit interactions are weaker, and the dots must be smaller due to a larger effective mass.

Bulk silicon has six equivalent conduction band minima located on the $\Delta$-lines, at $k_v \approx 0.84 k_0$ toward the six $X$ points of the Brillouin zone, where $k_0 = 2 \pi / a_0$ with $a_0 = 5.4\,\text{\AA}$ the lattice constant.\cite{Schaeffler1997:SST,Paul2004:SST,Dresselhaus2010:book} They are typically referred to as $\Delta$-valleys or $X$-valleys. In general, their degeneracy is lifted by strain, or by the presence of an interface.\cite{Ando1982:RMP,Paul2004:SST} In a (001)-grown silicon heterostructure, the four in-plane valleys are split by at least 10 meV from the two lower-lying $\pm z$ valleys, resulting in a twofold conduction band minimum. This remaining degeneracy is further split if the perpendicular confinement is asymmetric, resulting in an energy difference called the ground state gap.\cite{Ando1979:PRB,Weitz1996:SS,Goswami2007:NP,Saraiva2009:PRB,Culcer2010:PRB,Friesen2007:PRB,Friesen2010:PRB,Borselli2011b:APL} As the valley degeneracy is believed to be the main obstacle for silicon-based quantum computation\cite{Culcer2009:PRB,Culcer2010:PRB,Li2010:PRB}, a large valley splitting is desired. If this is the case, the multi-valley system can be reduced to an effective single valley qubit, a potentially nuclear-spin-free analog to the well-know GaAs counterpart.\cite{Culcer2009:PRB,Li2010:PRB} In fact, many recent experiments performed on Si/SiGe quantum dots have no evidence of valley degeneracy,\cite{Maune2012:N,Prance2012:PRL,Borselli2011:APL,Lai2011:SR,Hayes2008:CM,Simmons2011:PRL} indicating that the splitting is large enough to justify a single valley treatment. On the other hand, a recent proposal of valley-defined qubits uses the valley degree of freedom as a tool for gate operations.\cite{Culcer2012:PRL} This requires precise control of the ground-state gap, a challenging task for the future. In this work we assume that the valley splittings are larger than the typical energy scale of interest so that the effective single valley approximation is valid.

The spin relaxation and decoherence have been investigated theoretically and experimentally in silicon-based single and double dots from single to many electron occupancy.\cite{Simmons2011:PRL,Hayes2008:CM,Xiao2010:PRL,Pan2009:APL,Tyryshkin2006:PE,Tahan2002:PRB,Prada2008:PRB,Culcer2009:PRB,Glavin2003:PRB,Raith2011:PRB,Prance2012:PRL,Maune2012:N,Simmons2011:PRL,wang2011:JAP,Wang2010:PRB,shen2007:PRB,Sherman2005:PRB,Zwanenburg2012:CM} Our work completes these findings by a global, quantitative understanding of two-electron lateral silicon double quantum dots. We investigate the spin-orbit and hyperfine induced relaxation rate as a function of interdot coupling, detuning, and magnitude and orientation of the external magnetic field for zero and finite temperatures, and for natural and isotopically purified silicon. We pay special attention to the spin hot spots,\cite{Fabian1998:PRL} and investigate individual relaxation channels. This work is an extension to the findings in Ref.~\onlinecite{Raith2012:PRL} for GaAs, and we highlight the differences between these two materials. We fix the double dot orientation with respect to the crystallographic axes to that which is used most often in experiments. Our choices for other parameters are similarly guided by realistic values. Though we can not present results for the complete parametric space, exploring most direct experimental controls we expect the presented picture of double dot two electron spin relaxation in Si to be exhaustive, meaning the results listed below will remain qualitatively correct also beyond the specific parameter choices we make.

We find that due to the small spin-orbit coupling the spin relaxation rates are typically at least two orders of magnitude lower than in comparable GaAs dots, and that the relaxation rate peaks at spin hot spots are very narrow in parameter space. For detuned double dots, the energy spectrum close to the singlet-singlet anticrossing is qualitatively different from the GaAs counterpart, due to the rather small single-dot exchange coupling compared to the anticrossing energy. We also find that the hyperfine-induced relaxation rates of natural silicon are typically two and more orders of magnitude lower than the spin-orbit induced relaxation rates. The hyperfine-induced rates of purified silicon are further suppressed by about two orders of magnitude compared to natural silicon. Though the anomalous regime of nuclei dominating the relaxation, which we identified in GaAs\cite{Raith2012:PRL}, exists also in Si, here the different material parameters make it much harder to observe in practice. We therefore conclude that, concerning the relaxation, the nuclear field is negligible. Thus, the anisotropy of the spin-orbit field manifests in all relaxation rates we calculated, yielding the electrically controlled directional switch of the easy passage\cite{Stano2006:PRL} (a particular orientation of the magnetic field for which the relaxation as a function of some parameter is significantly lower than for other orientations), previously found in GaAs.\cite{Raith2012:PRL} A temperature of 0.1 K does not change our findings in any qualitative way.

\section{Model}

We consider a $\hat z = \left[001\right]$ grown top-gated Si/SiGe heterostructure defining a laterally coupled double quantum dot within the silicon layer with a fraction of $^{29}$Si isotopes. The double dot is charged with two electrons and not coupled to leads. Assuming the validity of the effective single valley approximation,\cite{Culcer2009:PRB} the Hamiltonian in the two-dimensional and the envelope function approximation reads
\begin{equation}\label{Model:Hamiltonian}
	H = \sum_{i=1,2} \left(T_i + V_i + H_{Z,i} + H_{\text{so},i} + H_{{\rm nuc},i}  \right) + H_{\text{C}} .
\end{equation}
The operators of position ${\bf r}$ and momentum ${\bf P}$ are two-dimensional, where $\hat x = \left[100\right]$ and $\hat y = \left[010\right]$. The single-electron terms are labeled by the electron index $i$. The kinetic energy is $T = \mathbf{P}^{2}/2m$, with the kinetic momentum $\mathbf{P} = -\text{i} \hbar \nabla + e \mathbf{A}$, the effective electron mass $m$, and the electron charge $-e$. For an external magnetic field, given by $\mathbf{B} = \left( B_{\parallel} \cos \gamma, B_{\parallel} \sin \gamma, B_z \right)$, where $\gamma$ is the angle between the in-plane component of $\mathbf{B}$ and $\hat x$, the vector potential in symmetric gauge reads $\mathbf{A} = B_z \left(-y,x\right)/2$. We neglect the orbital effects of the in-plane magnetic field component, which is a good approximation up to roughly 10 T for common heterostructures.\cite{epub8285} The electrostatic potential,
\begin{equation}\label{Model:Potential}
	V = \frac{\hbar^2}{2ml_{0}^4}\,\text{min}\{\left(\mathbf{r}-\mathbf{d}\right)^2\!\!,\left(\mathbf{r}+\mathbf{d}\right)^2\} + e\mathbf{E}\cdot\mathbf{r} ,
\end{equation}
consists of the biquadratic confinement\cite{Li2010:PRB,Pedersen2007:PRB} and the external electric field. For $\mathbf{E}=0$, the potential is minimal at $\pm\mathbf{d}$. The dimensionless ratio $2d/l_0$ will be in further called the interdot distance. The single dot scale is given by the confinement length $l_{0}$, and equivalently by the confinement energy $E_0 = \hbar^2/(ml_{0}^2)$. The electric field $\mathbf{E}$ is applied along the dot main axis $\mathbf{d}$, where the angle $\delta$ gives the in-plane orientation with respect to $\hat x$. Turning on $\mathbf{E}$ shifts the potential minima relative to each other by the detuning energy $\epsilon=2eEd$. The geometry is plotted in Fig.~1 of Ref.~\onlinecite{Stano2008:PRB}.

The Zeeman term is $H_Z = \left(g/2\right) \mu_{\text{B}} \boldsymbol{\sigma} \!\cdot\! \mathbf{B}$, with the vector of Pauli matrices $\boldsymbol{\sigma} = \left( \sigma_x, \sigma_y, \sigma_z \right)$, the effective Land\'{e} factor $g$, and the Bohr magneton $\mu_{\text{B}}$. The spin-orbit coupling, $H_{\text{so}} = H_{\text{br}} + H_{\text{d}}$, includes the Bychkov-Rashba\cite{bychkov1984:JPC,zutic2004:RMP} and the generalized Dresselhaus Hamiltonians,\cite{dresselhaus1955:PR,zutic2004:RMP,Golub2004:PRB,Nestoklon2008:PRB}
\begin{eqnarray}\label{Model:HamiltonianHso}
	H_{\text{br}} &=& \left(\hbar/2ml_{\text{br}}\right) \left(\sigma_x P_y - \sigma_y P_x \right) , \\
	H_{\text{d}} &=& \left(\hbar/2ml_{\text{d}}\right) \left(-\sigma_x P_x + \sigma_y P_y \right) ,
\end{eqnarray}
parameterized by the spin-orbit lengths $l_{\text{br}}$ and $l_{\text{d}}$. In this work we assume interface inversion asymmetry and choose $l_{\text{br}}$ and $l_{\text{d}}$ of comparable strength, according to Ref.~\onlinecite{Nestoklon2008:PRB}. The nuclear spins of $^{29}$Si dominantly couple through the Fermi contact interaction\cite{Assali2011:PRB,Khaetskii2003:PRB,schliemann2003:JPCM}
\begin{equation}\label{Model:HamiltonianHnuc}
H_{\rm nuc} = \beta \sum_n  {\bf I}_n \cdot \boldsymbol{\sigma}\, \delta ({\bf R} - {\bf R}_n),
\end{equation}
where $\beta$ is a constant, ${\bf I}_n$ is the spin of the $n$-th nucleus at the position ${\bf R}_n$, and $\mathbf{R} = \left( \mathbf{r}, z \right)$ is the three-dimensional electron position operator. Here we need to consider the finite extension of the wavefunction perpendicular to the heterostructure interface. We assume it is fixed to the ground state of a hard-wall confinement of width $w$. This defines the effective width,\cite{merkulov2002:PRB}
\begin{equation}\label{Model:effectivewidth}
h_{z} = \left[ \int \text{d}z \left| \psi(z)\right|^4 \right]^{-1} ,
\end{equation}
which evaluates to $h_{z} = 2w/3$. Finally, the Coulomb interaction is $H_{\text{C}} = e^2/4\pi\varepsilon \left|\mathbf{r}_1-\mathbf{r}_2\right|$, with the material dielectric constant $\varepsilon$.

The energy relaxation is enabled by phonons, whereas spin-orbit interactions allow for a spin-flip. In a (001)-grown quantum well of silicon, the electron-phonon coupling for intravalley scattering is the deformation potential of transverse acoustic (TA) and longitudinal acoustic (LA) phonons, given by\cite{Herring1956:PR,Hasegawa1960:PR,Prada2008:PRB,Wang2010:PRB,Pop2004:JAP,Duer1999:Nanotech}
\begin{equation}\label{Model:HamiltonianHelph}
H_{\text{ep}} = \text{i} \sum_{\mathbf{Q}, \lambda} \sqrt{\dfrac{\hbar Q}{2 \rho V c_{\lambda}}}D_{\mathbf{Q}}^\lambda \left[ b^{\dagger}_{\mathbf{Q},\lambda} \text{e}^{\text{i} \mathbf{Q} \cdot \mathbf{R}} \!-\! b_{\mathbf{Q},\lambda} \text{e}^{-\text{i} \mathbf{Q} \cdot \mathbf{R}} \right] \! ,
\end{equation}
where
\begin{equation}\label{Model:HamiltonianHelphV}
D_{\mathbf{Q}}^\lambda=(\Xi_{\text{d}} \mathbf{\hat e}_{\mathbf{Q}}^{\lambda} \cdot \mathbf{\hat Q} + \Xi_{\text{u}} \mathrm{\hat e}_{\mathbf{Q},z}^{\lambda} \hat Q_z).
\end{equation}
The phonon wave vector is $\mathbf{Q} = \left( \mathbf{q}, Q_z \right)$, and $\mathbf{\hat Q}=\mathbf{Q}/Q$. The polarizations are given by\cite{grodecka05a} $\lambda = \text{TA1, TA2, LA}$, the polarization unit vector reads $\mathbf{\hat e}$, and the phonon annihilation (creation) operator is denoted by $b$ ($b^{\dagger}$). The mass density,  the volume of the crystal, and the sound velocities are given by $\rho$, $V$, and $c_{\lambda}$, respectively. The efficiency of the electron-phonon coupling is set by the dilatation and shear potential constants, $\Xi_{\text{d}}$ and $\Xi_{\text{u}}$ respectively.

We define the relaxation rate (the inverse of the lifetime $T_1$) as the sum of the individual transition rates to all lower-lying states. Each rate (from $|i\rangle$ to $|j\rangle$) is evaluated using Fermi's Golden Rule at zero temperature,
\begin{equation}\label{Model:rate}
\Gamma_{ij} = \dfrac{\pi}{\rho V} \sum_{\mathbf{Q}, \lambda} \dfrac{Q}{c_{\lambda}} \left| D_{\mathbf{Q}}^\lambda \right|^2 \left| M_{ij} \right|^2 \delta(E_{ij} - E_{\mathbf{Q}}^\lambda) ,
\end{equation}
where $M_{ij} = \langle i | \text{e}^{\text{i} \mathbf{Q} \cdot (\mathbf{R}_1+\mathbf{R}_2)} | j \rangle$ is the matrix element of the states with energy difference $E_{ij}$, and $E_{\mathbf{Q}}^\lambda$ is the energy of a phonon with wave vector $\mathbf{Q}$ and polarization $\lambda$. In this work we focus on the singlet ($S$) and the three triplets ($T_-,T_0,T_+$) at the bottom of the energy spectrum.

Our numerical method is discussed in Refs.~\onlinecite{baruffa2010:PRL,*baruffa2010:PRB}. The extension to include the hyperfine coupling, Eq.~\eqref{Model:HamiltonianHnuc}, was introduced in Ref.~\onlinecite{Raith2012:PRL}. In this work, the two-electron basis for the configuration interaction method consists of 1156 Slater determinants, generated by 34 single-electron orbital states. The discretization grid is typically $135\times135$. The relative error for energies is below $10^{-5}$. The reliability of our code is confirmed by the evaluation of Eq.~\eqref{Model:rate} in an analytically solvable regime---weakly coupled dots in low magnetic fields. For details on this calculation, see the Appendix.

\begin{figure}
 \centering
 \includegraphics[width=0.85\linewidth]{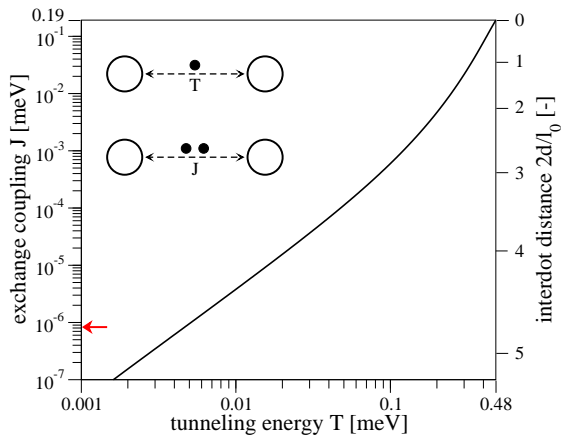}
\caption{(Color online) Calculated conversion between the single-electron tunneling energy $T$ (x axis), the two-electron exchange coupling $J$ (left y axis), and the interdot distance $2d/l_0$ (right y axis) neglecting nuclear spins. The arrow gives $E_{\text{nuc}}$, Eq.~\eqref{Results:Enuc}, of natural silicon.}
 \label{fig:TJ}
\end{figure}

We use the parameters of a SiGe/Si/SiGe quantum well grown along the $\hat z = \left[001\right]$ direction with a germanium concentration of 25\%. The two-dimensional electron gas is defined in the thin silicon layer with tensile strain.\cite{Schaeffler1997:SST,Yang2004:SST} The in-plane effective mass is isotropic, given by the transverse mass of the $X$ valley states,\cite{Ando1982:RMP} and we use $m=0.198m_e$,\cite{Rieger1993:PRB} where $m_e$ is the free electron mass. The effective Land\'e factor is $g=2$.\cite{Malissa2004:APL,Goswami2007:NP} Other material parameters read $c_{l}=9150$ m/s (for LA phonons), $c_{t}=5000$ m/s (for TA phonons), $\rho=2330$ kg/$m^3$, and $\varepsilon=11.9 \varepsilon_0$.\cite{LBGroupIII2002} The choice of deformation potential constants is not unique,\cite{Fischetti1996:JAP,Duer1999:Nanotech,Pop2004:JAP} and we use $\Xi_{\text{d}}=5$ eV and $\Xi_{\text{u}}=9$ eV according to Ref.~\onlinecite{LBGroupIII2002}. The hyperfine coupling parameter reads $\beta=-0.05\,\mu$eV nm$^3$, and $^{29}$Si has spin $I=1/2$. For natural silicon, the $^{29}$Si abundance is 4.7\%, and we use an abundance of 0.01\% for purified silicon. For the spin-orbit coupling strength we choose $l_{\text{br}}=38.5$ $\mu$m and $l_{\text{d}}=12.8$ $\mu$m.\cite{Malissa2004:APL,Nestoklon2008:PRB} The confinement length is $l_{0} = 20$ nm ($E_0 = 1.0$ meV), in line with realistic dot sizes.\cite{Hayes2008:CM,Gandolfo2005:JAP} The double dot is oriented as ${\bf d}$ $||$ [110]. The magnetic field is in-plane unless stated otherwise.

\begin{figure}
 \centering
 \includegraphics[width=0.99\linewidth]{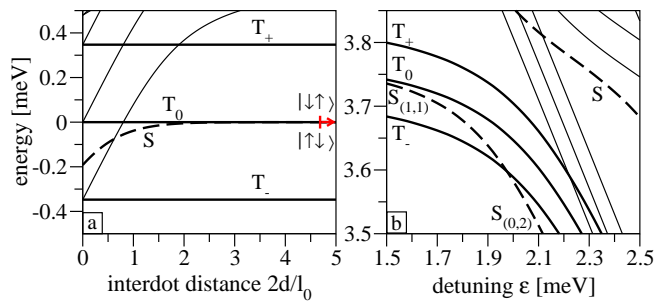}
\caption{(Color online) Calculated energies of the lowest states varying (a) the interdot coupling (at $B=3$ T), and (b) the detuning (at $B=0.5$ T, $2d/l_0=2.85$). Singlet states are given by dashed, triplets by solid lines. In (a), the energy of $T_0$ is subtracted, and in (b), the quadratic trend in $E$ is subtracted. The arrow in a) marks where $J=E_{\text{nuc}}$ (for natural silicon).}
 \label{fig:spectra}
\end{figure}

\begin{figure}
 \centering
 \includegraphics[width=0.75\linewidth]{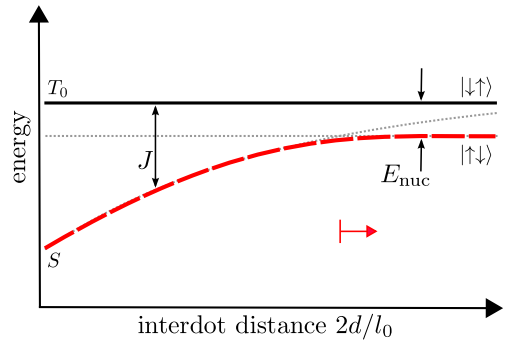}
\caption{(Color online) Schematic energy spectrum of an unbiased double dot showing the singlet S (dashed line) and the triplet $T_0$ (solid line). For large interdot distances, the exchange coupling $J$ is given by the hyperfine splitting $E_{\text{nuc}}$, Eq.~\eqref{Results:Enuc}, and the eigenstates change to $\left|\downarrow\uparrow\right>$ and $\left|\uparrow\downarrow\right>$.}
 \label{fig:schematic-spectrum_1}
\end{figure}

\section{Results: Unbiased Double Dot}
We parameterize in our model the coupling of the double quantum dot by the dimensionless interdot distance $2d/l_0$. The corresponding experimental observables are the tunneling energy $T$ (single-electron occupancy) and the exchange coupling $J$ (two-electron occupancy). The conversion between these three equivalent parameters is plotted in Fig.~\ref{fig:TJ} for clarity. 

The numerically calculated energy spectrum of the unbiased double dot is shown in Fig.~\ref{fig:spectra}. In this section, we choose a magnetic field of $B=3$ T. For the single dot ($d=0$) the exchange coupling, $J=E(T_0)-E(S)$, is $J_{\text{SD}} =$ 0.19 meV. The Zeeman energy, $E_{\text{Z}}=g \mu_{\text{B}} B$, exceeds $J$ for magnetic fields beyond 1.7 T. Consequently, we find in Fig.~\ref{fig:spectra}a that $T_-$ is the ground state for all interdot distances. The singlet therefore has an anticrossing with an excited triplet in the strong coupling regime, here at $J=75\,\mu$eV for our choice of parameters. This scenario is hardly met in comparable GaAs double quantum dots, because the required magnitude of the magnetic field is above 10 T. The silicon spectrum resembles the GaAs spectrum for magnetic fields below 1.7 T.

\begin{figure}
 \centering
 \includegraphics[width=0.75\linewidth]{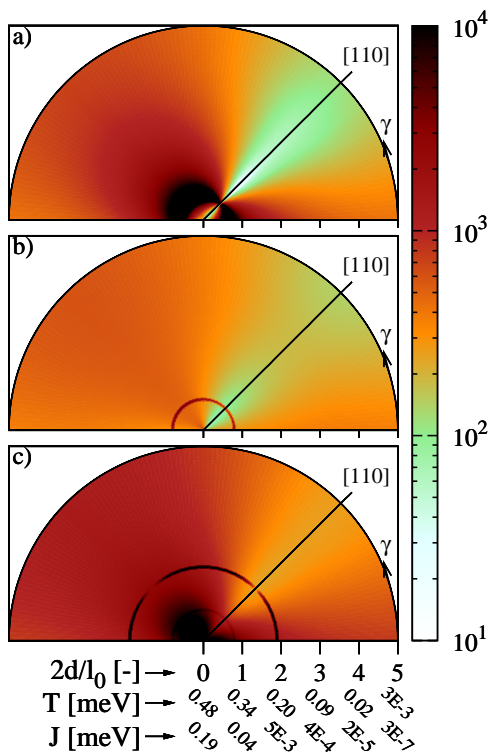}
\caption{(Color online) Calculated relaxation rates of (a) the singlet, (b) the triplet $T_0$, and (c) the triplet $T_+$ as a function of the in-plane magnetic field orientation $\gamma$ (angle) and the interdot distance $2d/l_0$ (radius of the polar plot), for a double dot at $B=$ 3 T.  The x and y axes correspond to crystallographic axes $\left[100\right]$ and $\left[010\right]$, respectively. The dot orientation ${\bf d}$ $||$ $\left[110\right]$ is marked by a line. The x axis is converted to the tunneling energy $T$ and the exchange $J$, in addition to $2d/l_0$. The rate is given in inverse seconds by the color scale. The system obeys $C_{2v}$ symmetry, so point reflection would complete the graphs.}
 \label{fig:relaxation_d}
\end{figure}

At large interdot distances, the hyperfine coupling induces a splitting of $S$ and $T_0$, given by
\begin{equation}\label{Results:Enuc}
E_{\text{nuc}} = 2\sqrt{ | \sum_{i=1,2} \langle \phi_a T_0 | H_{\text{nuc},i} | \phi_s S \rangle |^2} .
\end{equation}
In this regime, the lowest eigenstates are $\left|\uparrow\downarrow\right>=(S+T_0)/\surd{2}$ and $\left|\downarrow\uparrow\right>=(S-T_0)/\surd{2}$ (c.f.~Fig.~\ref{fig:schematic-spectrum_1}). We evaluate Eq.~\eqref{Results:Enuc} by averaging over random nuclear spin ensembles, and obtain $E_{\text{nuc}}\approx1$ neV for natural, and $E_{\text{nuc}}\approx0.04$ neV for purified silicon. This implies a crossover to the nuclear dominated regime at $2d/l_0 \gtrsim 4.7$ (red arrows in Figs.~\ref{fig:TJ}, \ref{fig:spectra}a, and \ref{fig:schematic-spectrum_1}) for natural, and at $2d/l_0 \gtrsim 5.4$ for purified silicon.

\begin{figure}
 \centering
 \includegraphics[width=0.9\linewidth]{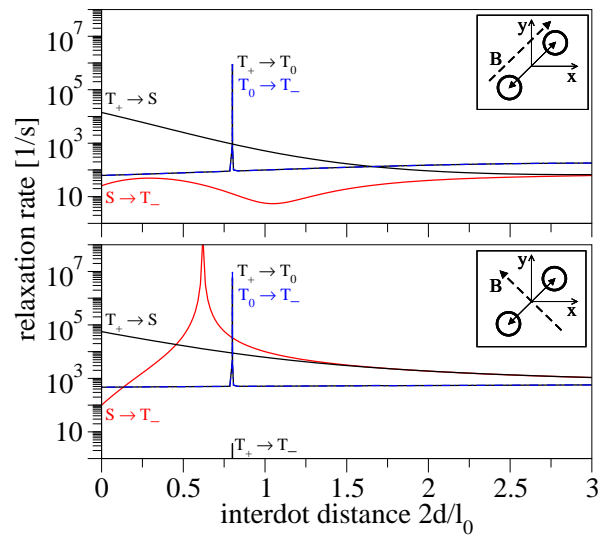}
\caption{(Color online) Calculated channel resolved relaxation rates vs.~interdot distance for both parallel (top) and perpendicular to $\mathbf{d}$ (bottom) in-plane magnetic field orientation ($B=3$ T). The relaxation channels of $T_+$ are black, of $T_0$ are blue, and of $S$ are red.}
 \label{fig:relaxation_channels}
\end{figure}

We plot the relaxation rates of the states $S$, $T_0$, and $T_+$, denoted in Fig.~\ref{fig:spectra}, as a function of the interdot distance and in-plane magnetic field orientation in Fig.~\ref{fig:relaxation_d}. We also give the relaxation rates of individual channels for the two principal axes, that is for the in-plane magnetic field component parallel and perpendicular to the dot main axis $\mathbf{d}$, in the upper and lower panel of Fig.~\ref{fig:relaxation_channels}, respectively. We find that the relaxation rate of the singlet is highly anisotropic,\cite{Golovach2004:PRL} which can be explained introducing an effective spin-orbit magnetic field (see below). The rates are minimal if $\mathbf{B} \parallel \mathbf{d}$, reaching the order of tens of milliseconds for any dot coupling strength (Fig.~\ref{fig:relaxation_channels}). We call this characteristic an easy passage.\cite{Stano2006:PRL,Stano2006:PRB} In the strong coupling regime, the rate away from the easy passage is enhanced by orders of magnitude. This results from the coupling of the singlet with the excited triplet, which favors the transition into $T_-$. For $\mathbf{B} \parallel \mathbf{d}$, the rate at the anticrossing is extremely sensitive to variations of $\gamma$, such that the easy passage becomes very narrow.

The relaxation rate of $T_0$ is given in Fig.~\ref{fig:relaxation_d}b. We find the same general anisotropic behavior, which is that the rate is minimal for $\mathbf{B} \parallel \mathbf{d}$. Figure \ref{fig:relaxation_channels} shows that the dominant channel of the relaxation is the transition $T_0 \rightarrow T_-$. Consequently, there is no impact from the singlet-triplet anticrossing. However, the anticrossing of the excited triplet with $T_0$ itself manifests in a very sharp peak of its rate. This spike is also anisotropic, with a difference of roughly one order of magnitude (not visible in Fig.~\ref{fig:relaxation_d}b due to its resolution).

Panel c) of Fig.~\ref{fig:relaxation_d} shows the relaxation rate of $T_+$. In addition to the anisotropic background, there are two spikes of enhanced rate generated by the anticrossings of $T_+$ with the excited triplets. The enhancement close to the single dot regime originates from the dominant $T_+ \rightarrow S$ transition (c.f.~Fig.~\ref{fig:relaxation_channels}). Interestingly, the anticrossing of the singlet hardly influences the overall trend of this relaxation channel. 

\begin{figure}
 \centering
 \includegraphics[width=0.75\linewidth]{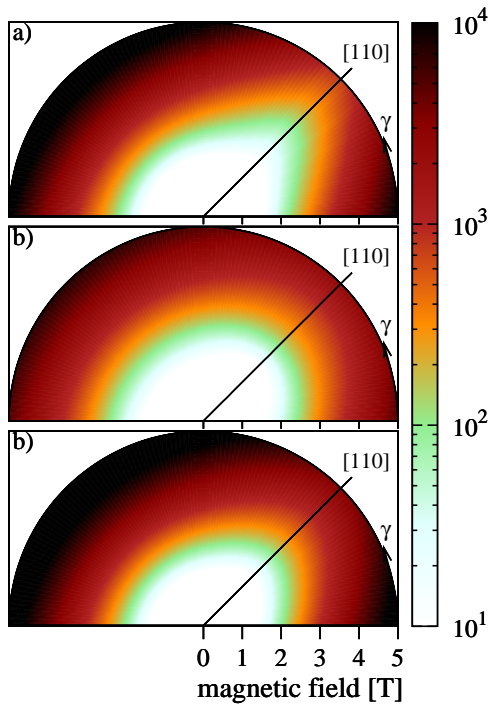}
\caption{(Color online) Calculated relaxation rates of (a) the singlet, (b) the triplet $T_0$, and (c) the triplet $T_+$ as a function of the in-plane magnetic field orientation $\gamma$ (angle) and the magnetic field magnitude (radius of the polar plot), for a double dot with $T$ = 0.1 meV. The layout with respect to the crystallographic axes is the same as in Fig.~\ref{fig:relaxation_d}. The rate is given in inverse seconds by the color scale.}
 \label{fig:relaxation_B}
\end{figure}

We plot in Fig.~\ref{fig:relaxation_B} the relaxation rates of a weakly coupled double dot as a function of in-plane magnetic field. Here we find the same qualitative behavior for all three panels. Similarly as in Fig.~\ref{fig:relaxation_d}, the relaxation rate is minimal for $\mathbf{B} \parallel \mathbf{d}$, but there are no spin hot-spots here.

\section{Results: Biased Double Dot}

\begin{figure}
 \centering
 \includegraphics[width=0.75\linewidth]{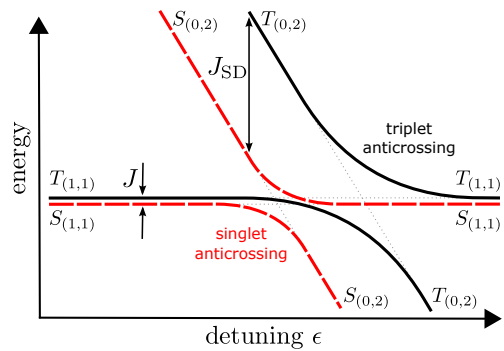}
\caption{(Color online) Schematic energy spectrum of a biased double dot without magnetic field. The singlets are given in dashed lines, the triplets in solid lines.}
 \label{fig:schematic-spectrum_2}
\end{figure}

In this section we consider a weakly coupled double dot with a finite detuning energy $\epsilon$. Figure \ref{fig:schematic-spectrum_2} introduces important characteristic energies in a schematic energy spectrum. The state charge character is given in brackets: (1,1) indicates that there is one electron in each dot, and (0,2) states that both electrons are in the same dot. In the spectra, we subtract the quadratic trend in the electric field $E$. This way, the (1,1) states are displayed horizontally unless influenced by anticrossings. The important quantities are the single-dot exchange coupling $J_{\text{SD}}$, the double-dot exchange coupling $J$, and the singlet and triplet anticrossing energy splittings, labeled in Fig.~\ref{fig:schematic-spectrum_2} respectively. The single-dot exchange coupling $J_{\text{SD}}$ is set by material parameters, i.e.~the Coulomb interaction, and system parameters, i.e.~the confinement length. For non-interacting electrons, $J_{\text{SD}}$ is equal to the confinement energy $E_0$, here 1 meV. For interacting particles, the Coulomb repulsion has strong impact on the symmetric ground state, the singlet, as here the electrons tend to group together. The first excited state, the triplet, is antisymmetric with respect to point reflection at the dot origin, and therefore less affected. As a consequence, \textit{$J_{\text{SD}}$ decreases as the Coulomb interaction strength increases}. For our choice of parameters $J_{\text{SD}} =$ 0.2 meV. In contrast, \textit{$J_{\text{SD}}$ increases as the confinement length decreases}. For instance, a confinement length of $l_0 =$ 17 nm results in $J_{\text{SD}} \approx$ 0.3 meV. This can be understood as follows.\cite{Merkt1991:PRB} On the one hand, a stronger confinement increases the Coulomb strength due to smaller effective particle distances $\left|\mathbf{r}_1-\mathbf{r}_2\right|$ in $H_{\text{C}}$. This is an effect somewhat linear in $l_{0}^{-1}$. Then, one could expect $J_{\text{SD}}$ to decrease. However, the confinement energy $E_0$ scales as $l_{0}^{-2}$, by what the exchange coupling increases in a similar way. This scaling dominates, such that the single-dot exchange coupling increases. The double-dot exchange coupling $J$ decreases exponentially with increasing interdot distance $d$.\cite{gorkov2003:PRB} In the weak coupling regime it holds $J \ll J_{\text{SD}}$, and we choose $d$ such that $J =$ 0.6 $\mu$eV. The anticrossing gap of a spin-alike pair of states at the (1,1) $\leftrightarrow$ (0,2) transition depends on the interdot distance as well. For increasing $d$ (decreasing $J$), these gaps decrease, that is the anticrossings vanish as $2d/l_0 \rightarrow \infty$.

The numerically calculated energy spectrum is plotted in Fig.~\ref{fig:spectra}b for a magnetic field of $B=0.5$ T. The spectrum is qualitatively different from the GaAs double dot counterpart (see Fig.~1 in Ref.~\onlinecite{Raith2012:PRL}). In a comparable GaAs double dot, the singlet and triplet anticrossings gaps are small compared to the single-dot exchange coupling. Consequently, the singlet anticrossing is well separated from the triplet anticrossing, and the excited singlet is close to $T_0$ in between these anticrossings.

\begin{figure}
 \centering
 \includegraphics[width=0.75\linewidth]{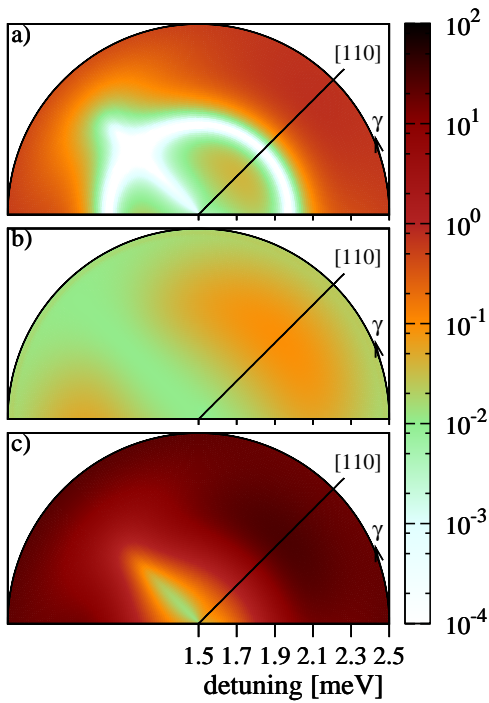}
\caption{(Color online) Calculated relaxation rates of (a) the first excited state ($S$ or $T_-$, see Fig.~\ref{fig:spectra}b), (b) $T_0$, and (c) $T_+$ as a function of the in-plane magnetic field orientation $\gamma$ (angle) and detuning energy (radius of the polar plot), for a double dot with $2d/l_0=2.85$ ($T$ = 0.1 meV), and $B=$ 0.5 T. The layout with respect to the crystallographic axes is the same as in Figs.~\ref{fig:relaxation_d} and \ref{fig:relaxation_B}. The rate is given in inverse seconds by the color scale.}
 \label{fig:relaxation_e}
\end{figure}

We plot the relaxation rates of the detuned double dot in Fig.~\ref{fig:relaxation_e}. Panel a) corresponds to the first excited state, that is $S$ for detuning energies up to 1.97 meV, and $T_-$ beyond. At the singlet-triplet anticrossing, the relaxation rate is very low as the transferred energy becomes very small. The easy passage occurs if the external, in-plane magnetic field is perpendicular to $\textbf{d}$. The same anisotropy is visible for the relaxation rates of $T_0$ and $T_+$, panel b) and c) respectively. There is no signature of the singlet-triplet anticrossing in the rate because of the exact compensation of individual relaxation channels. Also, there is no indication of the crossing of $T_+$ with the excited triplet $T_-$. This rate behavior---anisotropies, easy passage directional switch and the exact compensation---is analogous to a GaAs dot and we refer the reader to Ref.~\onlinecite{Raith2012:PRL} for a detailed discussion and explanation. Other anticrossings with excited triplets (at $\epsilon \approx 2.47$ meV) manifest in extremely narrow peaks of the rate, not visible in Fig.~\ref{fig:relaxation_e} at the current resolution.

\begin{figure}
 \centering
 \includegraphics[width=0.95\linewidth]{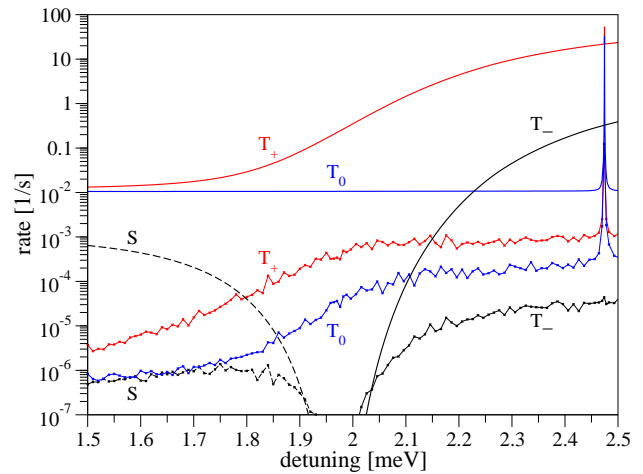}
\caption{(Color online) Calculated relaxation rates of a detuned double dot in an in-plane magnetic field ($B=$ 0.5 T, $\gamma = 3\pi/4$) as a function of detuning. The straight lines give the spin-orbit induced relaxation, the wiggly lines the hyperfine induced relaxation rates (natural silicon).}
 \label{fig:relaxation_detuning}
\end{figure}

Let us comment on the possible effects of nuclear spins. Comparing the interaction strengths with the spin-orbit fields, the former are expected to be negligible. Indeed, the Overhauser field characterizing the fluctuating collective nuclear field,\cite{schliemann2003:JPCM}
\begin{equation}\label{Results:Overhauserfield}
{\bf B}_{\text{nuc}} = \frac{\beta}{g\mu_{\text{B}}} \left< \sum_n {\bf I}_n \left| \psi({\bf R_n})\right|^2 \right> ,
\end{equation}
of natural silicon is of the order of tens of $\mu$T, for purified silicon even one order of magnitude lower. On the other hand, the effective spin-orbit field ($x_d, y_d=(x\pm y)/\sqrt{2}$),\cite{Stano2006:PRL}
\begin{equation}\label{Results:SOCfield}
{\bf B}_{\rm so}={\bf B} \! \times \! \{ x_d (l_{br}^{-1} - l_d^{-1}) [1\overline{1}0] + y_d (l_{br}^{-1} + l_d^{-1}) [110] \}/\sqrt{2} ,
\end{equation}
is about 2 mT at $B=1$ T. Still, in GaAs we have found that despite a similar discrepancy, there are cases where the nuclear field dominates the spin-orbit field, as the latter is quenched by symmetry imposed selection rules.\cite{Raith2012:PRL} Here, such a situation arises, in principle, too. However, due to the differences in material parameters, it requires an extremely weakly coupled double dot ($J$ of the order of sub peV; see App.~B for details), usually not pursued in experiments. The matrix elements of the spin flipping transitions are in Si therefore dominated by the spin-orbit fields, rather than nuclear spin fields and the same holds for anticrossing gaps. An illustration is given in Fig.~\ref{fig:relaxation_detuning}.

The second possibility we considered, was a dot detuned so far (so small $J$), that singlet and triplet $T_0$ become degenerate with respect to $E_{\rm nuc}$. The Hamiltonian eigenstates change, from entangled states into separable states with spin up or down in the left or right dot respectively (we show this schematically in Fig.~\ref{fig:schematic-spectrum_1}). Figures in Sec.~III cover this regime but the qualitative change in the eigenstate character has no visible effects on the relaxation rates (verified also for $2d/l_0 > 5$, not shown). This is because the relaxation to a fully spin polarized final state $T_+$ from the initial state $S$, $T_0$ (or any of their superposition, such as $\left|\uparrow \downarrow\right>$) proceeds through an individual single-dot spin-flip, with transition matrix element magnitude being essentially the same in all these cases.

Next we considered direct transitions due to random nuclear fields without phonon assistance. Such transitions are possible if the eigenstates have unsharp energies (finite lifetimes). As the states we are interested in are low lying, even at finite temperature their energy broadening is so small that the resulting nuclear induced spin relaxation is negligible.

Finally, we considered the consequences of the random character of the nuclear field, which blurs the electron energies. This statistical, rather than quantum mechanical, uncertainty can be grasped roughly by convoluting the relaxation curves with a Gaussian with an appropriate width, depending on which parameter we change, defined ultimately by the energy $g \mu_{\text{B}}{\bf B}_{\text{nuc}}$. We find this width to be unnoticeable small---as an example, the extremely narrow peaks in Fig.~5 survive practically untouched by such smoothening. We therefore conclude that unpolarized nuclear spins in natural or purified Si are not expected to be visible in the electron spin relaxation within the parametric space we investigate. We find that such a situation might occur only for very small external fields ($B\leq 0.01$ T) or very weakly coupled dots ($J\leq$ peV).

The figures presented and results discussed in this article were for zero temperature. In our model, a finite temperature amounts solely to allowing for energy increasing transitions (phonon absorption), in addition to phonon emission processes only which are present at zero temperature. We analyzed this possibility, adopting a typical experimental value of 100 mK. We have not found any case where such additional transitions would change the relaxation rates in any significant way (figures not shown). Our conclusion from these investigations is that the relaxation character, most notably its anisotropies, will not be influenced by experimentally relevant sub-Kelvin temperatures.

\acknowledgments
This work was supported by DFG under grant SPP 1285 and SFB 689. P.S. acknowledges support by meta-QUTE ITMS NFP 26240120022, CE SAS QUTE, EU Project Q-essence, APVV-0646-10 and SCIEX.

\appendix
\section{Analytical Calculation of Relaxation Rates}
In this section we analytically calculate the relaxation rate, Eq.~\eqref{Model:rate}, adopting several approximations. The calculations prove useful to explain the physical mechanism and to verify our numerical results. The validity of the approximations will be discussed afterwards. In the following, the hyperfine coupling is neglected.

Approximating the sum in Eq.~\eqref{Model:rate} by an integral, and rewriting the $\delta$-function with respect to the $z$ component of $\mathbf{Q}$, we obtain ($i \neq j$)
\begin{eqnarray}\label{Appendix:rate}
\Gamma_{ij} =& \dfrac{E_{ij}}{8 \pi^2 \rho \hbar^2} \sum_{\lambda} \int\!\text{d}\mathbf{q} \int\!\text{d}Q_z \, \dfrac{Q}{c_{\lambda}^3 \bar{Q}_{z}^\lambda} \left| D_{\mathbf{Q}}^\lambda \right|^2 \left| M_{ij} \right|^2 \nonumber \\
&\times \left[\delta(Q_z - \bar{Q}_{z}^\lambda) + \delta(Q_z + \bar{Q}_{z}^\lambda)\right] ,
\end{eqnarray}
where $\bar{Q}_{z}^\lambda = \sqrt{E_{ij}^2/(\hbar^2 c_{\lambda}^2) - q^2}$. Assuming the validity of the dipole approximation, the matrix element reads
\begin{equation}\label{Appendix:matrix_element_dipole_approx}
M_{ij} \approx \text{i} \langle i | \mathbf{q} \cdot (\mathbf{r}_1 + \mathbf{r}_2) | j \rangle,
\end{equation}
where $| i \rangle$ and $| j \rangle$ are the spin-orbit coupled two electron eigenstates. Note that the contribution of the wave function overlap along the $z$ direction in $M_{ij}$ is about 1,\cite{Stano2006:PRB} which is consistent with the two-dimensional approximation. We restrict ourselves to weakly coupled double dots, i.e.~$d \gg l_0$, and incorporate the effect of spin-orbit coupling perturbatively via a Schrieffer-Wolff transformation.\cite{lowdin1951:JCP,Schrieffer1966:PR,Winkler2003:book} The eigenstates then read ($l$ labels the electrons)
\begin{equation}\label{Appendix:eigenstates}
| i \rangle = \text{e}^{-\text{i}O}\left( | i \rangle_0 + \sum_k \sum_{l=1,2} \frac{_0\langle k | \bar{H}_{\text{so},l} | i \rangle_0}{E_{i}^0 - E_{k}^0} | k \rangle_0 \right) ,
\end{equation}
with the transformation operator $O = - \sum_l \mathbf{n}_{\text{so},l} \cdot \boldsymbol{\sigma_l}/2$, where
\begin{equation}\label{Appendix:nso}
\mathbf{n}_{\text{so},l} = \left( \frac{x_l}{l_{\text{d}}} - \frac{y_l}{l_{\text{br}}}, \frac{x_l}{l_{\text{br}}} - \frac{y_l}{l_{\text{d}}}, 0\right) ,
\end{equation}
and the effective spin-orbit operator $\bar{H}_{\text{so},l} = \bar{H}_{\text{so},l}^Z + \bar{H}_{\text{so},l}^{(2)}$, where
\begin{eqnarray}
\bar{H}_{\text{so},l}^Z &=& \frac{g}{2}\mu_{\text{B}}(\mathbf{n}_{\text{so},l} \times \mathbf{B})\cdot\boldsymbol{\sigma}_l , \label{Appendix:Hso_1}\\
\bar{H}_{\text{so},l}^{(2)} &=& \frac{\hbar}{4m} \left(\frac{1}{l_{\text{d}}^2} - \frac{1}{l_{\text{br}}^2}\right) L_{z,l} \sigma_{z,l} + \text{const} . \label{Appendix:Hso_2}
\end{eqnarray}
Here, $L_z = l_z + (e/2) r^2 B_z$, where $l_z$ is the operator of angular momentum. The states in Eq.~\eqref{Appendix:eigenstates} labeled with subscript $0$ are eigenstates of the Hamiltonian
\begin{equation}\label{Appendix:Hamiltonian}
	H^0 = \sum_{i=1,2} \left(T_i + V_i + H_{Z,i} \right) + H_{\text{C}},
\end{equation}
their eigenenergies are denoted as $E^0$. We use the Heitler-London ansatz\cite{Heitler1927:ZfP} to approximate the eigenstates of Eq.~\eqref{Appendix:Hamiltonian}.

We use Eq.~\eqref{Appendix:eigenstates} to evaluate the matrix element $M_{ij}$. It is straightforward to show that contributions from coupling within the lowest four-dimensional subspace, $\mathcal{M}=\{S,T_-,T_0,T_+\}$, are zero or exponentially suppressed in $d/l_0$. As a result, the relaxation requires coupling via higher states. Neglecting the $L_z$ contribution to the effective spin-orbit coupling, Eq.~\eqref{Appendix:Hso_2}, we obtain
\begin{widetext}
\begin{equation}\label{Appendix:matrix_element}
\frac{M_{ij}}{\text{i}g\mu_{\text{B}}} = \sum_{k\notin\mathcal{M}} \sum_{l=1,2} \left[ \frac{_0\langle i | (\mathbf{n}_{\text{so},l} \times \mathbf{B})\cdot\boldsymbol{\sigma}_l | k \rangle_0}{E_{i}^0 - E_{k}^0} {}_0\langle k | q_x x_1 + q_y y_1 | j \rangle_0 + \frac{_0\langle k | (\mathbf{n}_{\text{so},l} \times \mathbf{B})\cdot\boldsymbol{\sigma}_l | j \rangle_0}{E_{j}^0 - E_{k}^0} {}_0\langle i | q_x x_1 + q_y y_1 | k \rangle_0\right] .
\end{equation}
\end{widetext}

The singlet is symmetric with respect to the inversion operator $I$ (point reflection in real space), the triplets are antisymmetric.\cite{baruffa2010:PRL} Consequently, it follows from Eq.~\eqref{Appendix:matrix_element} that, within the dipole approximation, the singlet-triplet transition is forbidden. We also find that Eq.~\eqref{Appendix:matrix_element} forbids a $T_+ \leftrightarrow T_-$ transition because the effective spin-orbit operator, $\bar{H}_{\text{so},l}^Z$, acts on only one of the two electron spins. Let us now look at the transition between $T_0$ and $T_\pm$.

To evaluate Eq.~\eqref{Appendix:matrix_element}, we reduce the infinite sum over $k$ to cover only states within the energy window of about the confinement energy, $\hbar^2/(ml_{0}^2)$. Additionally, we can exclude any singlet from the sum, because the electron-phonon operator does not act in spin space. What is left can be captured by the Heitler-London approach.

Let $| \text{R}0 \rangle$ be the (orbital) ground state of a single dot shifted to the ``right'' by $d$, i.e.~the Fock-Darwin state of the right dot with the principal quantum number $n=0$ and the orbital quantum number $l=0$. Analogously we define the ground state of the ``left'' dot. The  properly symmetrized triplet lowest in energy is
\begin{equation}\label{Appendix:HL-triplet}
| \Psi_T \rangle_0 = \left( | \text{R}0,\text{L}0 \rangle - | \text{L}0,\text{R}0 \rangle \right) \otimes | T \rangle  / \sqrt{2} .
\end{equation}
The orbitally excited triplets can be constructed analogously, using $| \text{R}1 \rangle$, and $| \text{L}1 \rangle$, the displaced Fock-Darwin states with $n=0$ and $|l|=1$:
\begin{eqnarray}\label{Appendix:HL-k}
| k_\pm \rangle_0 =&\left( | \text{R}0,\text{L}1 \rangle - | \text{L}1,\text{R}0 \rangle \pm ( | \text{R}1,\text{L}0 \rangle - | \text{L}0,\text{R}1 \rangle )\right) \nonumber \\
 & \otimes | T \rangle / 2 .
\end{eqnarray}

Neglecting the wave function overlap of states localized in different quantum dots, we calculate the matrix elements $_0\langle \Psi_T | x_1 | k_\pm \rangle_0$ and $_0\langle \Psi_T | y_1 | k_\pm \rangle_0$ analytically, yielding
\begin{eqnarray}
_0\langle \Psi_T | x_1 | k_+ \rangle_0 &=& l_0 /\sqrt{8} , \label{Appendix:matrix_elements_1}\\
_0\langle \Psi_T | y_1 | k_+ \rangle_0 &=& {\rm sgn}(l) \text{i} l_0 /\sqrt{8} , \label{Appendix:matrix_elements_2}
\end{eqnarray}
and $_0\langle \Psi_T | x_1 | k_- \rangle_0 = {}_0\langle \Psi_T | y_1 | k_- \rangle_0  =0$. We use Eqs.~\eqref{Appendix:matrix_elements_1} and \eqref{Appendix:matrix_elements_2} as an approximation for the matrix elements in Eq.~\eqref{Appendix:matrix_element}. We also require the matrix elements of Pauli matrices respecting the spin quantization axis along $\mathbf{B}$. They read
\begin{eqnarray}\label{Appendix:spin_matrix_element}
& \langle T_\pm |  \boldsymbol{\sigma}_1 | T_0 \rangle = \nonumber \\
& = \frac{\text{e}^{\mp\text{i}\gamma}}{2\sqrt{2}} \begin{pmatrix} \cos(\gamma-\theta)+\cos(\gamma+\theta)\pm2\text{i}\sin(\gamma) \\ \sin(\gamma-\theta)+\sin(\gamma+\theta)\mp2\text{i}\cos(\gamma) \\2\sin(\theta) \end{pmatrix},
\end{eqnarray}
where $\theta=\cos^{-1}(B_z/B_\parallel)$. The energy differences in Eq.~\eqref{Appendix:matrix_element} are approximated by the confinement energy, $\hbar^2/(m l_{0}^2)$.

With these ingredients, we can solve Eq.~\eqref{Appendix:rate}, integrating over the phonon momentum, and obtain
\begin{eqnarray}\label{Appendix:rate2}
&\Gamma_{T_0 \rightarrow T_-} = \Gamma_{T_+ \rightarrow T_0} = \frac{m^2 l_{0}^8 \mathcal{L}_{\text{so}}^{-2}}{12 \pi \rho \hbar^{10}} \left(g \mu_{\text{B}} B\right)^7 \nonumber \\
& \times \left[c_{l}^{-7} \left( \frac{3}{35} \Xi_{\text{u}}^2 + \frac{2}{5} \Xi_{\text{u}} \Xi_{\text{d}} + \Xi_{\text{d}}^2 \right) + c_{t}^{-7} \frac{4}{35} \Xi_{\text{u}}^2 \right] ,
\end{eqnarray}
with the effective spin-orbit length $\mathcal{L}_{\text{so}}$, defined by
\begin{equation}\label{Appendix:effective_SOlength}
\mathcal{L}_{\text{so}}^{-2} = \begin{cases} 2 \left( l_{\text{br}}^{-2} + l_{\text{d}}^{-2} \right) & \text{if $\theta = 0$,} \\
l_{\text{br}}^{-2} + l_{\text{d}}^{-2} - 2 \frac{\sin(2\gamma)}{l_{\text{br}} l_{\text{d}}} & \text{if $\theta = \pi/2$.} \end{cases}
\end{equation}

Now we discuss the validity of the approximations used during the derivation of Eq.~\eqref{Appendix:rate2}. The matrix element $M_{ij}$ is calculated using the dipole approximation, Eq.~\eqref{Appendix:matrix_element_dipole_approx}. It requires that the energy difference between the transition states, here $T_0$ and $T_\pm$, fulfills $E_{ij} \ll \hbar c_\lambda / l_0$.\cite{Stano2006:PRB} Using $E_{ij} = g \mu_{\text{B}} B$, and $c_\lambda = c_t$, we obtain the condition $B \ll 1.4$ T. We consider also a weakly coupled double dot, $d \gg l_0$, to comply with most experiments. This limit ensures negligible matrix elements among the states of $\mathcal{M}$, and justifies the Heitler-London approximation. Here, the spectrum also develops bundles of eigenenergies separated by the confinement energy $\hbar^2/(ml_{0}^2)$, a fact used to approximate the energy differences in Eq.~\eqref{Appendix:matrix_element}. Note that within the restriction of the dipole approximation ($B \ll 1.4$ T), the Zeeman energy ($E_{\text{Z}}\ll 0.16$ meV) is negligible compared to the confinement energy ($E_0 = 1$ meV). The Schrieffer-Wolff transformation is the essential tool for a perturbative treatment of spin-orbit coupling in the double dot.\cite{Stano2005:PRB} Perturbation theory with the unitarily transformed Hamiltonian yields results which are higher order in small quantities compared to the original Hamiltonian.\cite{Stano2005:PRB,Golovach2004:PRL} Finally, we note that, since $L_z$ is symmetric with respect to the inversion $I$, the perturbation $\bar{H}_{\text{so},l}^{(2)}$, Eq.~\eqref{Appendix:Hso_2}, vanishes for all transitions $T_0 \leftrightarrow T_\pm$.

\begin{figure}
 \centering
 \includegraphics[width=0.90\linewidth]{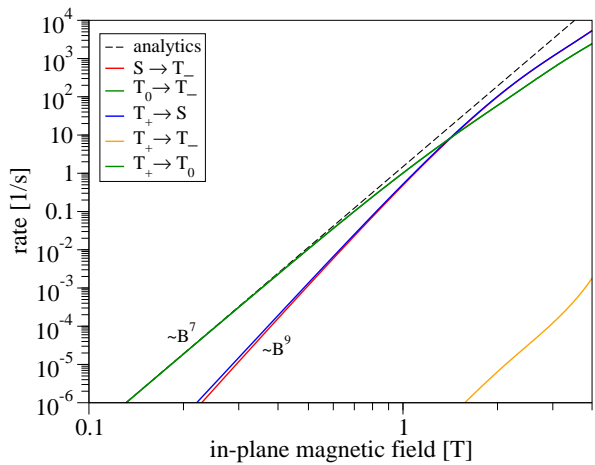}
\caption{(Color online) Calculated relaxation rates of individual transition channels as a function of in-plane magnetic field for a weakly coupled quantum dot. The magnetic field is oriented along $\left[\bar110\right]$ ($\gamma = 3\pi/4$), and the dots along $\left[110\right]$ ($\delta = \pi/4$). The interdot distance is $2d/l_0 =$ 2.85, yielding the tunneling energy $T =$ 0.1 meV. The dashed, black line gives the analytical relaxation rate, evaluated with Eq.~\eqref{Appendix:rate2}.}
 \label{fig:analytics}
\end{figure}

We compare the analytical formula for the relaxation rate of the transition $T_0 \rightarrow T_-$, and $T_+ \rightarrow T_0$, given in Eq.~\eqref{Appendix:rate2}, with the numerical results in Fig.~\ref{fig:analytics}. We find perfect agreement for low magnetic fields, in line with the condition $B \ll 1.4$ T. For larger magnetic fields, the results significantly deviate from the $B^7$ power law, due to the break down of the dipole approximation. We also find that the $S \leftrightarrow T_\pm$ relaxation channels, which we found to be zero in the lowest order dipole approximation due to their symmetry, show $B^9$ dependence, indicating that the relaxation is driven by the second order term of $\mathbf{q}$. Being of higher order, the relaxation rate, for small $\mathbf{B}$, is at least one order of magnitude lower than the $T_0 \leftrightarrow T_\pm$ transitions. 

\section{Nuclear dominance}

Here we estimate the parameters at which, for the electron spin relaxation, nuclear spins dominate the spin-orbit fields. Comparing the strengths of the two effective fields, as done in the main text, one does not expect such a situation to arise, unless at very small (below ten millitesla or so) external magnetic fields. This regime is not usually met in experiments, where a sizeable Zeeman splitting is necessary for electron spin manipulations and measurements. We have found in our previous work on GaAs quantum dots\cite{Raith2012:PRL} that despite the discrepancy, there are anomalous cases where the above expectation fails and nuclei are indeed the dominant channel. This happens in a weakly coupled double dot biased to the $S_{1,1}-S_{0,2}$ anticrossing, if the corresponding anticrossing gap, $E_{S-S}$, is small enough. Namely, due to the absence of the spin-orbit coupling between states $T_0$ and $S$, the small magnitude of the nuclear induced wavefunction admixture is compensated by the small energetic distance of the two states. The very same mechanism is also present in Si, raising the question for its conditions to become manifest.

We will illustrate the case by comparing Si to GaAs. For this, we assume that the single dot energy $E_0$ and the Zeeman energy are the same in the two quantum dots, each built in one of the two materials. We estimate the ratio of the exchange energies (which characterize the interdot coupling) below for which the nuclei dominate. As described above, this happens if
\begin{equation}
g \mu_B B_{\rm nuc}/E_{S-T} \gtrsim g \mu_B B_{\rm so} / E_{T-T},
\label{eq:Jcrit}
\end{equation}  
where the effective magnetic fields are defined in Eqs.~\eqref{Results:Overhauserfield} and Eq.~\eqref{Results:SOCfield}, $E_{S-T}$ is the difference of the energy of the states $S$ and $T_0$, approximately equal to the $S_{1,1}-S_{0,2}$ anticrossing gap, and $E_{T-T}$ is the energy difference of the state $T_0$ and the closest excited triplet, which we approximate by the orbital energy scale $E_0$. We define a ``critical'' $E_{S-T}$ energy difference by Eq.~\eqref{eq:Jcrit} with an equality sign. Approximating the two electron wavefunctions by Slater determinants composed of localized Fock-Darwin states of a single dot, we get the following auxiliary results, valid for large interdot distances,
\begin{equation}
E_{S-S} \approx \frac{e^2}{\sqrt{2} \epsilon_0 \epsilon_r d} \exp(-d^2/l_0^2),
\end{equation}
and 
\begin{equation}
J \approx \exp(-2 d^2/l_0^2) \frac{4d}{\sqrt{\pi}l_0} \frac{\hbar^2}{m l_0^2}.
\end{equation}
Both of these quantities fall off exponentially with the interdot distance in weakly coupled dots. However the $S-S$ anticrossing gap scales as the tunneling energy $T$, whereas the exchange energy is much smaller, $J \sim T^2/U$ (here $U$ is the charging energy).\cite{burkard1999:PRB} 
With these we get for the ratio of critical exchange energies,
\begin{equation}
\frac{J_{\rm crit}^{\rm Si}}{J_{\rm crit}^{\rm GaAs}} \sim \left( p\frac{(I \beta l_{\rm so} \epsilon_r)_{\rm Si}}{(I \beta l_{\rm so} \epsilon_r)_{\rm GaAs}} \right)^2 \approx 10^{-6}. 
\label{eq:Jratio}
\end{equation}
Here $p$ is the fraction of the isotope $^{29}$Si. We have found previously that in a GaAs quantum dot with parameters typical in experiments, the nuclei dominance requires exchange energies of the order of 0.1$\mu$eV. As follows from Eq.~\eqref{eq:Jratio}, in silicon the requirements are much more stringent and thus less suitable for such an effect demonstration. The reason for this are different material parameters, most importantly much weaker coupling of the conduction electrons to the nuclear spins and low fraction of atoms with non-zero nuclear magnetic moment in silicon.

\end{document}